**Effect of thermal history on magnetism in UCoGa**

P. Opletal[1], P. Proschek[1], B. Vondráčková[1], D. Aurélio[1], V. Sechovský[1] and J. Prokleška[1]

[1]*Charles University, Faculty of Mathematics and Physics, Ke Karlovu 3, Praha 2 121 16, Czech Republic*

**Abstract**

Single crystals of UCoGa have been grown in different conditions and subsequently annealed in order to provide a collection of samples representing various quality as to concentration of lattice defects. The different sample quality "grades" have been characterized by values of residual electrical resistivity. Correlations of magnetic parameters (coercive field, Curie temperature) with residual resistivity have been determined and domain-wall pinning by crystal defects has been confirmed as the underlying mechanism of coercive field in strongly anisotropic ferromagnets.

**Introduction**

The ferromagnetic U*TX* (*T* = late transition d-electron metal, *X* = p-electron element) compounds, which crystallize in the hexagonal ZrNiAl-type structure (space group P-62m) and exhibit strong uniaxial magnetocrystalline anisotropy [1], have recently been going through a renaissance of interest due to the possible presence of quantum phase transitions (QPT) connected with the loss of magnetism[2–5].

This group of compounds is known to be homogeneous, both from the point of view of crystal structure and ground-state magnetic ordering, yet diverse as to size of magnetic moment magnitudes and ordering temperatures [1]. The defined and smooth scaling of magnetic properties within one structure type allows studies related to the presence of QPT and related phenomena – it has been shown theoretically [6] and observed experimentally [7] that the temperature of the tricritical point and the critical field, scale with the magnitude of the saturated moment in the presence of the first order QPT. The presence of the first order QPT is conditioned by sufficiently low disorder (usually crystal defects in sufficiently clean materials), however, disorder is a property difficult to grasp experimentally, much less quantifiable exactly.

Various crystal-lattice defects may be considered as the main source of disorder. The pinning of very narrow domain walls by crystal defects has been suggested to be responsible for the large coercive fields observed in UPtAl and UIrAl single crystals [8,9]. However, no connection to the crystal growth process nor the possible relation to presence of crystal lattice imperfections or other types of disorder has been discussed.

In this paper we show a clear connection between the technology of preparation, presence of disorder and the low temperature magnetic properties of one hexagonal U*TX* ferromagnet, namely UCoGa. We have prepared two single crystalline samples, which subsequently underwent different thermal treatments. Our results document that the thermal history influences both the disorder (as indicated by the residual resistivity) and magnetism of a crystal.

**Experimental**

Polycrystalline precursors for UCoGa single-crystal growth have been obtained by melting the UCo$_2$ and UGa$_2$ master-alloy ingots together. High purity (U SSE purified, Co 3N5, Ga 7N) starting element metals were used. Two percent of gallium was added to compensate for high Ga evaporation during precursor melting and single crystal growth. The single crystals have been grown from melted precursors by Czochralski method using a triarc furnace with a pulling rate of 10 mm/hr, crucible clockwise rotation 2.5 rpm and seed counter-clockwise rotation of 4 rpm for the first single crystal and 0 rpm for the second one.

For annealing, the single crystals were wrapped in a tantalum foil and sealed in a quartz tube in vacuum at $10^{-6}$ mbar. Structure and composition of the crystals was checked by X-ray Laue method (Laue diffractometer of Photonic Science) and EDX analysis (scanning electron microscope Tescan Mira I LMH). Samples were cut for specific experiments by a wire saw using the solution of distilled water, glycerin and silicon carbide powder. In order to properly investigate the effect of annealing, a differential scanning calorimetry (DSC) was performed on a piece of the polycrystalline ingot using SETSYS Evolution, resulting in the determination of the melting temperature as $T_M = 930$ °C.

Magnetization measurements were performed on oriented crystals in a MPMS 7 XL SQUID magnetometer (Quantum Design) with the magnetic field applied along the c-axis, the easy magnetization direction. Demagnetization field was not subtracted from our data and consequently absolute values should be compared only for a given sample. The electrical resistivity was measured in PPMS (Quantum Design) using four-probe technique with current applied within the basal plane. The heat capacity was measured in PPMS (Quantum Design) using relaxation method.

**Results**

We have successfully grown two single crystals of the UCoGa compound (further denoted as SC1 and SC2). In order to achieve controllable difference in the preparation, the rotation of the seed was turned off during growth of SC2. From both crystals, plate-like samples were cut (perpendicular to the crystallographic c-axis). Magnetic and electrical transport measurements were done on as-grown and annealed crystals, in order to investigate their properties and the effects of the thermal history in detail.
The results obtained on samples grown in different conditions and subjected to different thermal treatments are reviewed in Table 1. Two pieces of SC2 (from the part of the crystal closer to the seed) were annealed for 3 weeks at 800°C and 900°C (denoted as SC2B800 and SC2A900), respectively and then cooled at rate of 2 K/hour to room temperature. As there was no difference in investigated properties between these two samples, we will only present and discuss results obtained on the SC2A900 sample further on. The physical properties of the third piece of SC2 (denoted as SC2C, piece taken from a part of the crystal located far from seed and close to melt) have not been influenced by the annealing process. Then we show results obtained on the as grown crystal only.
Based on results obtained from annealing SC2, we chose to anneal SC1 samples for 3 weeks at 800°C (SC1 A800) so as to prevent possible interactions with the tantalum foil.
In all cases the annealing does not lead to any observable changes to the crystallinity or composition of the samples, at least as observed within the resolution of the used methods.

The thermomagnetic curves measured during cooling in a magnetic field of 0.01 T applied along the c-axis (Fig 1) were used for first characterization. The temperature of the inflection point of the $M(T)$ curve has been taken as a reasonable estimate of Curie temperature ($T_C$). The SC2C sample which shows a considerably lower $T_C$ value and much smoother variation of the magnetization across the ordering temperature. The estimated $T_C$ values for all studied samples are listed in Table 1. The measured temperature dependence of the heat capacity shows a second order phase transition independently on the ordering temperature as expected for the itinerant ferromagnetic material (see Fig S1).

The magnetization curves measured at 1.8K (Fig. 2) manifest the evolution of coercive field ($H_c$). We can see that the lowest $H_c$ value was observed for the annealed sample SC2A900, whereas the as-grown SC2C shows the highest coercive field. The as-grown SC1A crystal exhibits an $H_c$ value comparable to that reported in literature [10]. When we inspect the temperature evolution of the magnetization curves an exponential decrease of $H_c$ with increasing temperature is found for all investigated crystals (Fig 2). This is similar behavior as described for soft ferromagnets in [11]. The coercive field is proportional to domain wall pinning factor $k$ [12], which is exponentially dependent on temperature [11]

$k(T) = k(0)*\exp(-T/(\beta T_C))$     (1),

where $T_C$ is Curie temperature and $\beta$ is critical exponent for pinning constant.
Results of the fits within the pinning model for soft ferromagnets are given in Table 1. The different $H_c$ values corresponding to different samples can be attributed to the different density of defects reflecting different thermal histories of the samples. The defects act as pinning centers which prevent smooth movement of domain walls. It has been shown that the coercive field is proportional to square root of the density of defects and that it also depends on the domain wall area [13]. Since our single crystals are of the nearly the same chemical composition (within the sensitivity of EDX) and $T_C$ values, sample invariable values of the exchange and magnetic anisotropic energy in all samples may be expected, which should result in comparable domain-wall areas in all our samples. Consequently, we can determine the relative change in the density of defects of SC1A and SC2A taking the SC2A900 sample as a reference with lowest coercive field (see Table 1).
The exponential decrease of coercive field is also found in amorphous metals and magnetic garnets with uniaxial anisotropy. In these materials the coercive field originates from an uniaxial anisotropy field [14]. In our case the uniaxial anisotropy field is much larger than in garnets, which leads to higher coercive fields. We cannot exclude that the both contribute to coercivity in our case. But by comparison we see that $k(0)$, which depends on density of pinning sites and energy of pinning sites and is proportional to $H_C(0)$, scales with the quality of crystals.

The residual electrical resistivity as a result of scattering of conduction electrons on defects and other lattice imperfections is widely considered as a probe of material quality as concerns to lattice defects. Therefore measurements of the electrical resistivity at low temperatures were performed on the same samples as those used for the magnetization measurements. The temperature dependences of electrical resistivity of all studied samples are qualitatively comparable (see Fig. 4) – a monotonous decrease from room temperature down to Curie temperature and a sudden increase of the rate of decrease at $T_C$ with decreasing temperature. This is caused by the suppression of the spin-disorder contribution due to ferromagnetic ordering. At the lowest temperatures the resistivity asymptotically approaches the residual resistivity value $\rho_0$. A notable exception is the SC2C sample with the resistivity dominated by the large residual resistivity in the entire temperature range (the residual resistivity ratio RRR=R(100K)/R(2K)=1.8), contrary to all other samples (RRR=6.3-11.4) and the effect of magnetic ordering on the character of the temperature dependence is negligible. The high degree of disorder in the SC2C sample is reflected also in a considerably reduced $T_C$ value in comparison to all the other samples. The observed $\rho_0$ values for all samples are also shown in Table 1. There we can see that the lowest residual resistivity and coercive force were obtained for the SC2A900 sample whereas the highest $\rho_0$ and $H_c$ values refer to for SC2C.

**Discussion and Conclusions**

The considerable difference in quality between the as-grown SC1 and SC2 crystals can be attributed to different parameters of the growth procedure. In contrast to the SC1 crystal rotation, the SC2 crystal was not rotating during the growth that probably caused less optimized solidification process in the latter case presumably causing a higher concentration of defects. The defect concentration was possibly enhanced by a slight off-stoichiometry of a sample taken more distant from the seed, cf. the SC2A and SC2C crystals. The increasing off-stoichiometry with the distance from the seed could be explained by gallium deficiency increasing with time due to continuous gallium evaporation from the melt. The off-stoichiometry is, however, indeed slight, well below the lower detection limit of EDX equipment.

Annealing at 900°C, which is just a few degrees below the melting temperature, significantly improved the quality of the SC2A single crystal to the levels fully comparable with optimally grown SC1, as documented by both the observed magnetization and electrical resistivity behavior. The SC2C sample represent an exceptional case where the damage of the lattice assisted by off-stoichiometry were far from optimal and probably new local thermal equilibrium has been reached, which could not be influenced by annealing. This is also seen in change of critical exponent $β$, which differs from other crystals. If we compare the residual resistivity of the SC2A900 sample with those available in literature of other members of the U$TX$ family with the ZrNiAl crystal structure, we see that the best values are of the order of 10 μΩ.cm (URhAl [15], UCoAl [3,16], UNiGa [17]). These values are fully comparable to the best values obtained in this study.

The high density of defects may explain the somehow peculiar shape of the thermomagnetic curves as measured at low fields - there is a notable shape difference between the $M(T)$ curves measured on the SC2C sample and the others. The SC2C thermomagnetic curve resembles Brillouin-like behavior whereas other temperatures dependencies show a pronounced shoulder in the vicinity of the Curie temperature. This behavior is suppressed by larger field (see Fig S2) and is not unique within the given family of U$TX$ compounds. It has been observed in e.g. URhAl [15], although the reports are scarce presumably due to lack of measurements performed in low fields. Our explanation is based on the correlation between the sample quality and the observed phenomena. Based on the technology and the sample analysis we expect that majority of the structure defects to be low dimensional, namely point defects (vacancies) and line defects (dislocations).

At zero magnetic field a domain wall will be located in the position minimizing the system energy, in this case the sum of the energy of the domain wall itself and the involved domains. Consequently the narrow 180 degree domain walls (due to the strong uniaxial anisotropy) will tend to embrace the defect (or be in the place of minimum microstress in the case of finite size defects) (see e.g. [18]). As discussed earlier this leads to higher observed defect density in the case of SC2C sample, consequently to higher density of domain walls. Particularly in this system (U$TX$) with strong uniaxial anisotropy (and therefore very narrow domain walls) this may explain the lower ordering temperatures as a result of energetically less stable configuration even in the cases where the off-stoichiometry is not noticeable.

In small fields as used in this study, the domain walls are virtually immobile and only possible mechanism for the change of the bulk magnetization of the sample is the change of polarity ('switching') of the magnetization of individual domains. In the case of SC2C sample the typical size of domains is expected to be relatively smaller which is leading to easier 'switching' of the domain moment and consequently to the Brillouin like behavior of magnetization below the ordering temperature. On the other hand other samples with (on average) larger domains remain in the 'as ordered' state several K below the ordering temperature, when the $μ_BB/k_BT$ parameter is sufficiently large to start 'switching' the (on average larger then in previous case) domain's moment and in result mimics the Brillouin like behavior of sample's magnetization. However, a detail local probe (MFM, submicrometer Hall magnetometry) capable of observing magnetization of individual domains would be necessary to analyze such processes and possibly confirm proposed scenario.

High coercive-field values related to domain-wall pinning have been reported in other UTX ferromagnets from the ZrNiAl family (UPtAl, UIrAl). The authors reported the residual resistivity values of single crystals of UPtAl $\rho_0 = 68$ μΩ.cm [8] and UIrAl $\rho_0 = 87$ μΩ.cm [19]. No annealing treatment has been mentioned, however. Nevertheless, the reported high $\rho_0$ values corroborate the scenario that the crystal defects were the primary source of the reported high coercive fields similar to our case.

Our study focused on UCoGa highlights the importance of the technology of single crystal preparation, and the subsequent thermal treatment in UTX compounds with ZrNiAl crystal structure, which can considerably influence the physical properties. This is of special interest in the context of phenomena investigation on the verge of magnetism, where the sample quality may act as a scenario-changing parameter. We also believe that conclusions of our study may be extended to more materials far beyond the limited family of UTX ferromagnets.

**Acknowledgment**
This work is a part of the research program GACR 16-06422S, which is financed by the Czech Science Foundation. D.A. acknowledges support of the European Regional Development Fund; OP RDE; Project: CARAT (No. CZ.02.1.01/0.0/0.0/16_026/0008382). Experiments were performed in the Materials Growth and Measurement Laboratory MGML LM2018096 (see: http://mgml.eu).

Table 1: Summary of the fits of experimental data to the pinning model for soft ferromagnets and density of defects in relation to density defects of SC2A900 $\sigma_0$ ($T_C$ - Curie temperature, $H_{coer}$ - coercive magnetic field, $\beta$ – critical exponent for pinning constant, $\rho_0$ – residual resistivity, $\sigma$ – density of defects).

| UCoGa |  | $T_c$ [K] | $H_{coer}$ (0) [T] | $\beta$ | $\rho_0$ [μΩ.cm] | $\sigma$ [$\sigma_0$] |
|---|---|---|---|---|---|---|
| SC2A900 | annealed | 49 | 0.04 | 0.24 | 14.4 | 1.00 |
| SC1A800 | annealed | 48 | 0.07 | 0.22 | 19.4 | 2.70 |
| SC1A | as-grown | 48 | 0.07 | 0.25 | 18.5 | 2.70 |
| SC2A | as-grown | 48 | 0.08 | 0.24 | 44.8 | 3.54 |
| SC2C | as-grown | 44 | 0.19 | 0.33 | 75.0 | 21.12 |

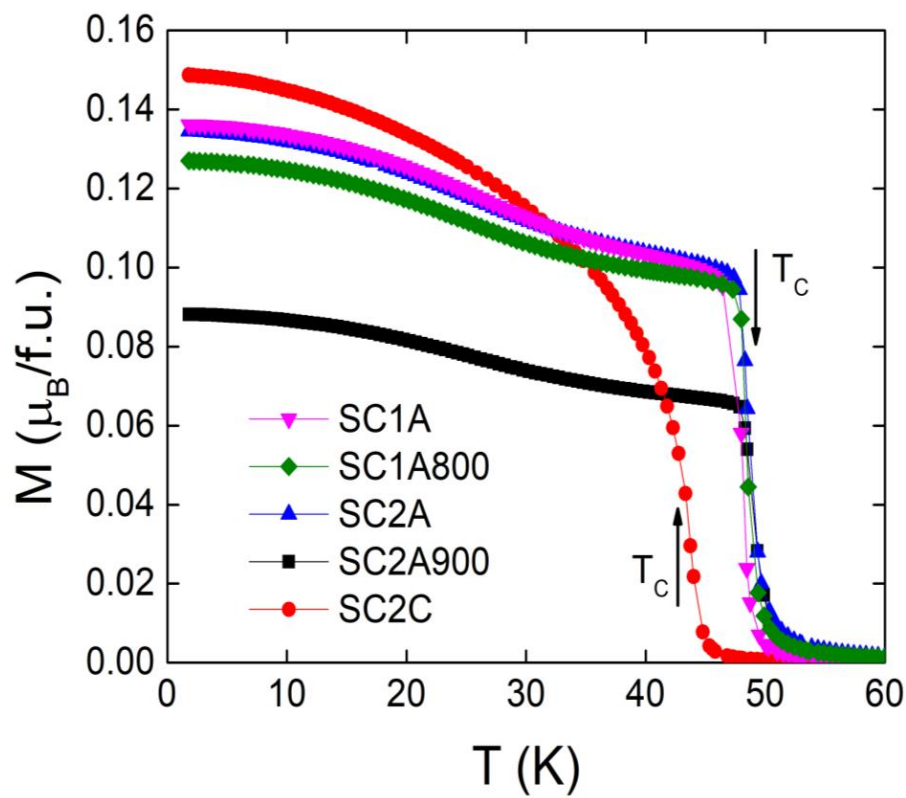

Figure 1: Temperature dependence of magnetization of studied crystals in the magnetic field of 0.01 T applied along the c-axis.

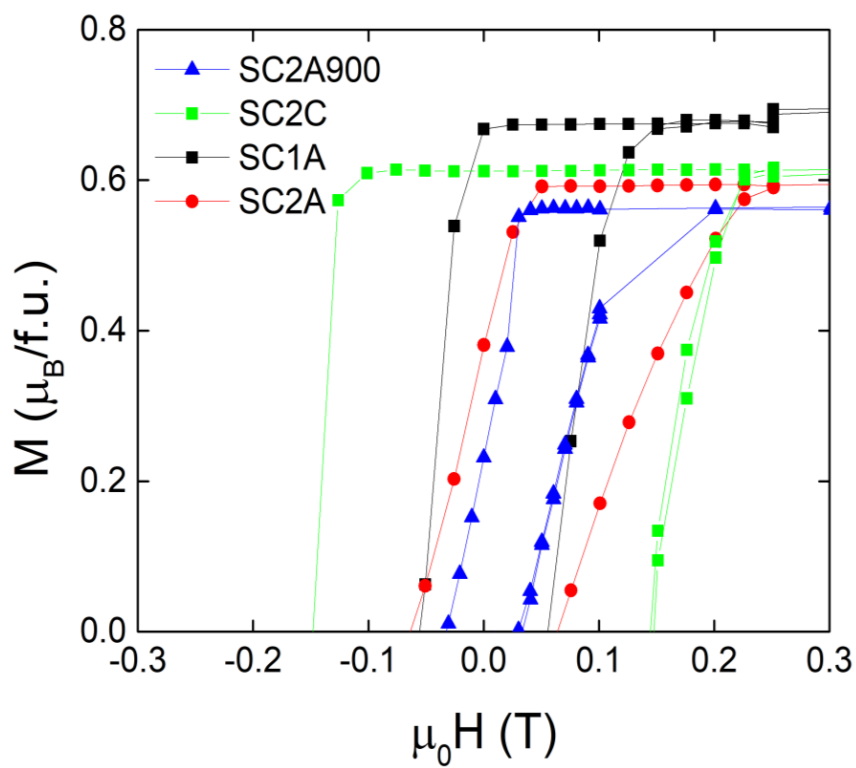

Figure 2: Magnetization curves of studied crystals measured at 1.8K in the magnetic field applied along the c-

axis.

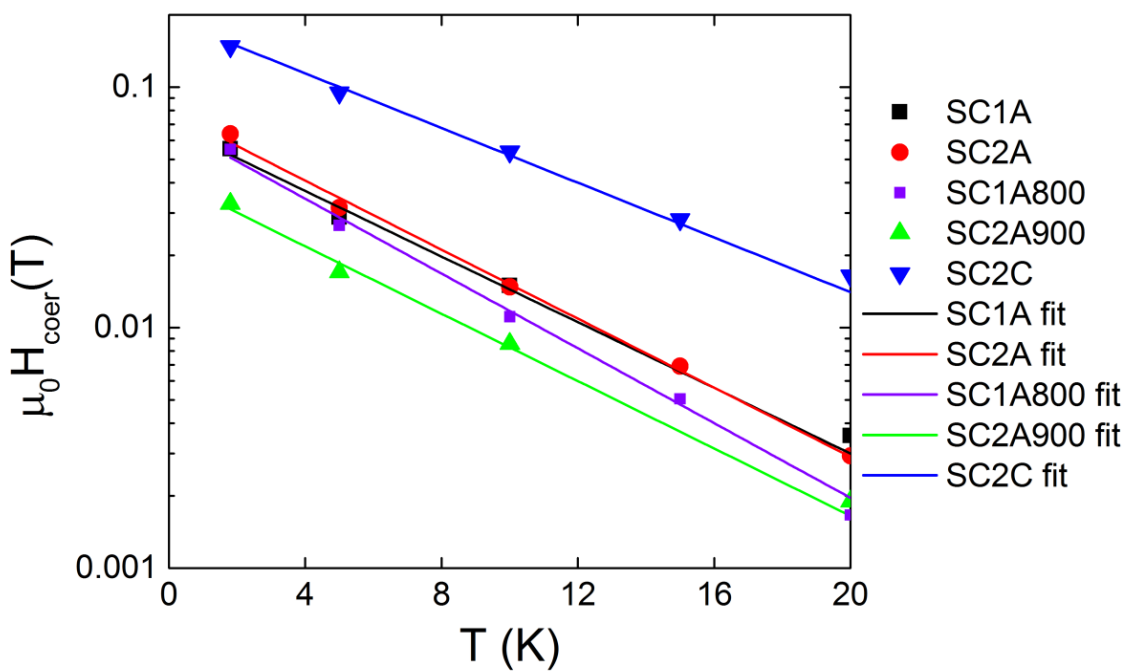

Figure 3: Temperature dependence of coercive field of studied crystals. Logarithmic scale (vertical axis) is used to emphasize the exponential behavior following equation (1). The lines represent the fits (see detail in text).

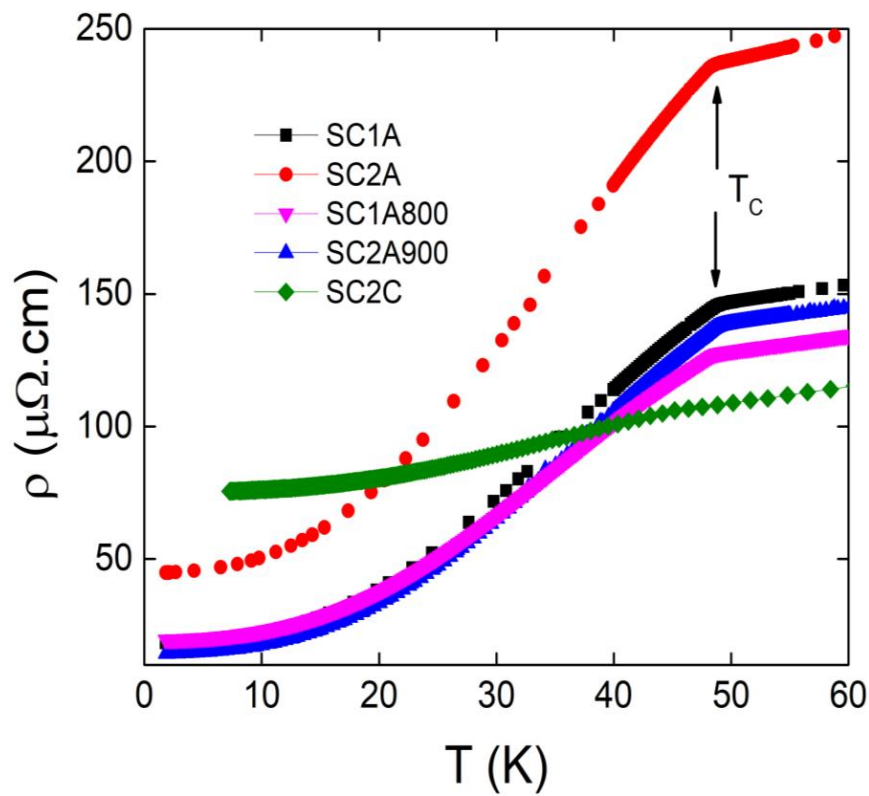

Figure 4: Temperature dependence of electrical resistivity of different samples with current applied perpendicular to the c-axis.

[1] V. Sechovsky, L. Havela, Chapter 1 Magnetism of ternary intermetallic compounds of uranium, in: K.H.J. Buschow (Ed.), Handb. Magn. Mater., Elsevier, 1998: pp. 1–289.

[2] Y. Shimizu, D. Braithwaite, B. Salce, T. Combier, D. Aoki, E.N. Hering, S.M. Ramos, J. Flouquet, Unusual strong spin-fluctuation effects around the critical pressure of the itinerant Ising-type ferromagnet URhAl, Phys. Rev. B. 91 (2015) 125115. doi:10.1103/PhysRevB.91.125115.

[3] N. Kimura, N. Kabeya, H. Aoki, K. Ohyama, M. Maeda, H. Fujii, M. Kogure, T. Asai, T. Komatsubara, T. Yamamura, I. Satoh, Quantum critical point and unusual phase diagram in the itinerant-electron metamagnet UCoAl, Phys. Rev. B. 92 (2015) 35106. http://link.aps.org/doi/10.1103/PhysRevB.92.035106.

[4] D. Aoki, T. Combier, V. Taufour, T.D. Matsuda, G. Knebel, H. Kotegawa, J. Flouquet, Ferromagnetic Quantum Critical Endpoint in UCoAl, J. Phys. Soc. Japan. 80 (2011) 94711. doi:10.1143/JPSJ.80.094711.

[5] M. Míšek, J. Prokleška, P. Opletal, P. Proschek, J. Kaštil, J. Kamarád, V. Sechovský, Pressure-induced quantum phase transition in the itinerant ferromagnet UCoGa, AIP Adv. 7 (2017). doi:10.1063/1.4976300.

[6] D. Belitz, T.R. Kirkpatrick, J. Rollbühler, Tricritical Behavior in Itinerant Quantum Ferromagnets, Phys. Rev. Lett. 94 (2005) 247205. doi:10.1103/PhysRevLett.94.247205.

[7] M. Brando, D. Belitz, F.M. Grosche, T.R. Kirkpatrick, Metallic quantum ferromagnets, Rev. Mod. Phys. 88 (2016) 25006. http://link.aps.org/doi/10.1103/RevModPhys.88.025006.

[8] A.V. Andreev, Y. Shiokawa, M. Tomida, Y. Homma, V. Sechovský, N.V. Mushnikov, T. Goto, Magnetic Properties of a UPtAl Single Crystal, J. Phys. Soc. Japan. 68 (1999) 2426–2432. doi:10.1143/jpsj.68.2426.

[9] A. V Andreev, Magnetization study of a UIrAl single crystal, J. Alloys Compd. 336 (2002) 77–80. doi:http://dx.doi.org/10.1016/S0925-8388(01)01889-8.

[10] H. Nakotte, F.R. de Boer, L. Havela, P. Svoboda, V. Sechovsky, Y. Kergadallan, J.C. Spirlet, J. Rebizant, MAGNETIC-ANISOTROPY OF UCoGa, J. Appl. Phys. 73 (1993) 6554–6556.

[11] A. Raghunathan, Y. Melikhov, J.E. Snyder, D.C. Jiles, Theoretical Model of Temperature Dependence of Hysteresis Based on Mean Field Theory, IEEE Trans. Magn. 46 (2010) 1507–1510. doi:10.1109/TMAG.2010.2045351.

[12] D. Jiles, Introduction to magnetism and magnetic materials, n.d.

[13] H.R. Hilzinger, H. Kronmüller, Statistical theory of the pinning of Bloch walls by randomly distributed defects, J. Magn. Magn. Mater. 2 (1975) 11–17. doi:http://dx.doi.org/10.1016/0304-8853(75)90098-0.

[14] G. Vértesy, I. Tomáš, Z. Vértesy, On the temperature dependence of domain wall pinning field in soft, uniaxial magnetic materials, J. Phys. D. Appl. Phys. 35 (2002) 625–630. doi:10.1088/0022-3727/35/7/310.

[15] T. Combier, Ferromagnetic quantum criticality in the uranium-based ternary compounds URhSi, URhAl, and UCoAl, Université de Grenoble, 2014.

[16] P. Opletal, J. Prokleška, J. Valenta, P. Proschek, V. Tkáč, R. Tarasenko, M. Běhounková, Š. Matoušková, M.M. Abd-Elmeguid, V. Sechovský, Quantum ferromagnet in the proximity of the tricritical point, Npj Quantum Mater. 2 (2017). doi:10.1038/s41535-017-0035-6.

[17] L. Jirman, V. Sechovský, L. Havela, W. Ye, T. Takabatake, H. Fujii, T. Suzuki, T. Fujita, E. Brück, F.R. De Boer, Magnetic and transport properties of UNiGa, J. Magn. Magn. Mater. 104–107 (1992) 19–20. doi:https://doi.org/10.1016/0304-8853(92)90683-F.

[18] S. Chikazumi, Physics of ferromagnetism, Oxford University Press, 1997.

[19] A. V Andreev, N. V Mushnikov, F. Honda, V. Sechovský, P. Javorský, T. Goto, Magnetic, magnetoelastic and other electronic properties of a UIrAl single crystal, J. Magn. Magn. Mater. 272–276, S (2004) E337–E339. doi:http://dx.doi.org/10.1016/j.jmmm.2003.11.273.

**Supplementary info**

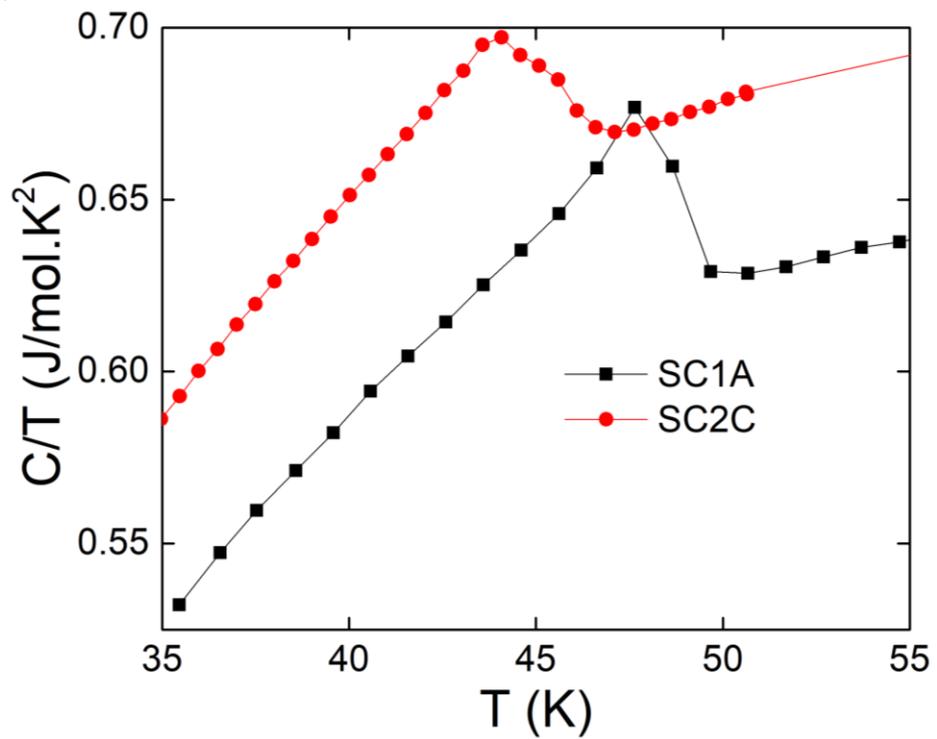

Figure S1: Temperature dependence of heat capacity of two different sample of UCoGa plotted as C/T vs T.

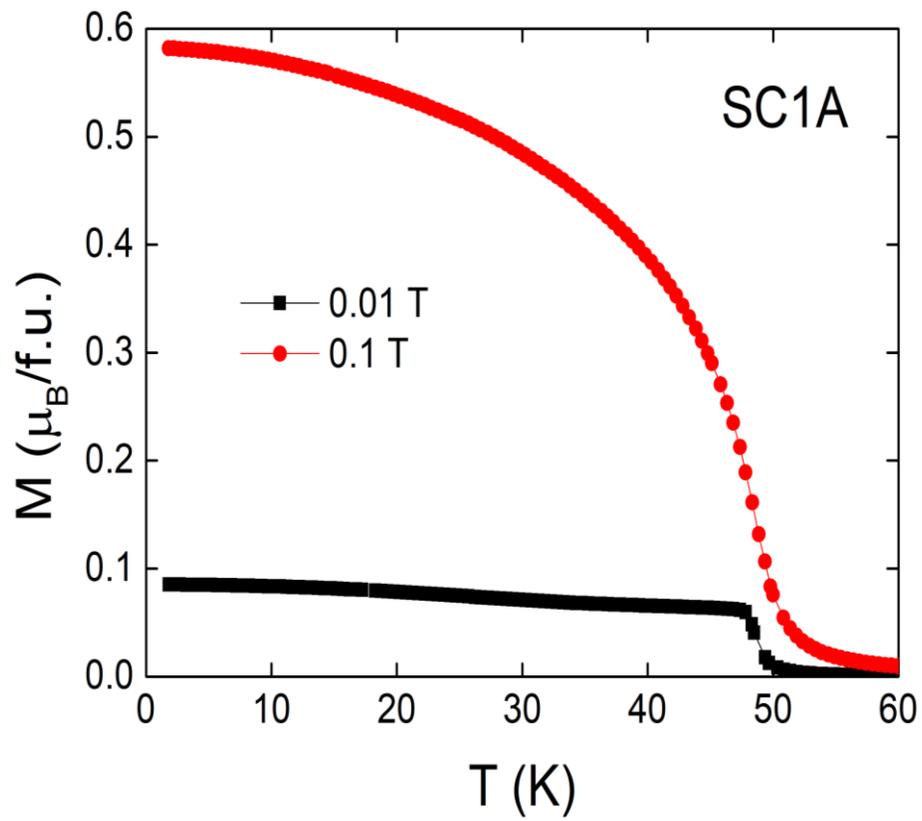

Figure S2: Temperature dependence of magnetization of sample SC1A of UCoGa in different applied magnetic field.